\documentclass[10pt,twocolumn,twoside]{elsart}

\usepackage{graphicx}
\usepackage{amsmath}

\begin{document}

\begin{frontmatter}

\title{Three-Particle Correlations from Parton Cascades in Au+Au Collisions}

\author{G. L. Ma$^{a}$ } \author{  Y. G.
Ma$^{a}$} \ead{ygma@sinap.ac.cn. Corresponding author },
\author{S. Zhang$^{a,b}$, X. Z. Cai$^{a}$, J. H. Chen$^{a,b}$,  Z. J. He$^a$, }
\author{H. Z. Huang$^c$, J. L. Long$^a$, W. Q. Shen$^{a}$, X. H. Shi$^{a,b}$, C. Zhong$^a$, J. X. Zuo$^{a,b}$ }
\author{ }

\address{$^a$ Shanghai Institute of Applied Physics, Chinese Academy of
Sciences, Shanghai 201800, China\\
$^b$ Graduate School of the Chinese Academy of Sciences, Beijing
100080, China\\ $^c$ Department of Physics and Astrophysics,
University of California, Los Angeles, CA90095, USA}

\date{\today}

\begin{abstract}

We present a study of three-particle correlations among a trigger
particle and two associated particles in Au + Au collisions at
$\sqrt{s_{NN}}$ = 200 GeV using a multi-phase transport model
(AMPT) with both partonic and hadronic interactions. We found that
three-particle correlation densities in different angular
directions with respect to the triggered particle (`center',
`cone', `deflected', `near' and `near-away') increase with the
number of participants. The ratio of `deflected' to `cone' density
approaches to 1.0 with the increasing of number of participants,
which indicates that  partonic Mach-like shock waves can be
produced by strong parton cascades in central Au+Au collisions.

\vspace{1pc}
\end{abstract}


\end{frontmatter}

\begin{description}
   \item[I. Introduction ]
\end{description}

Ultra-relativistic heavy ion collisions may provide conditions
sufficient for the formation of a deconfined plasma of quarks and
gluons~\cite{QCD}. Experimental results from RHIC indicate that a
strongly-interacting partonic matter (termed sQGP) has been
created in the early stage of central Au + Au collisions at
$\sqrt{s_{NN}}$ = 200 GeV at RHIC~\cite{White-papers}. Jet-like
azimuthal correlation is one of the important hard probes to
explore the natures of the newly formed matter. The
disappearance~\cite{jet-ex1} and re-appearance~\cite{jet-ex2} of
back-to-back high transverse momentum ($p_T$) particles from jets
have been proved to result from the interactions between
jet-partons and the hot and dense medium created in central Au+Au
collisions. Recently, an interesting splitting of the away side
peak has been observed in the di-hadron azimuthal angle
($\Delta\phi$) correlation distribution between soft associated
particles and high $p_{T}$ trigger particles in central Au + Au
collisions at
RHIC~\cite{sideward-peak1,sideward-peak2,sideward-peak3}. Such a
double peak structure on the away-side is consistent with
preferential conic emission of particles from jets and/or
shock-wave induced collective motion from jet-medium interactions.
We will refer to the observed double peaks on the away-side as the
Mach-like structure without necessary implication on the dynamical
mechanism.

Several theoretical interpretations about the Mach-like structure
have been proposed. For instances, St\"ocker et al. proposed the
Mach-like structure from jets traversing the dense medium as a
probe of the equation of state (EOS) and the speed of sound in the
medium \cite{Stocker}. Casalderrey-Solana and Shuryak et al.
argued a shock wave generation because jets travel faster than the
sound in the medium \cite{Casalderrey}. They fitted the broad
splitting structure on the away side in di-hadron azimuthal
correlation with a Mach-cone shock wave mechanism. Vitev has shown
that the cancellation of collinear bremsstrahlung in QCD medium
can lead to large angle emission of gluons~\cite{Vitev}. Koch and
Wang et al. used a Cherenkov radiation model with negative
dispersion relation to produce the Mach-like structure
\cite{Koch}. In Ref.~\cite{Armesto}, Armesto proposed that the
medium-induced gluon radiation could be affected by the collective
flow in the medium. It has also been argued by M\"uller et al.
that a Mach-like structure can appear via the excitation of
collective plasmon waves by moving color charges associated with
the leading jet~\cite{Ruppert}. Renk and Ruppert found that in
order to reproduce the experimental data a large fraction (about
$90\%$) of the lost energy of jet has to be channelled to excite a
shock wave in a dense medium at a soft point of
EOS~\cite{Thorsten}. Satarov et al. investigated Mach shocks
induced by partonic jets in expanding quark-gluon
plasma~\cite{Sat}. However, Chaudhuri and Heinz reported no
observation of Mach-like structures in di-hadron $\Delta\phi$
correlations from jet quenching dynamically in a hydrodynamic QGP
fluid~\cite{Chaudhuri}. A consistent dynamical picture for the
generation of the Mach-like structure in particle correlations has
yet to emerge and further investigations are needed.

In order to shed light on the puzzle of the dynamical origin of
the splitting structure on the away-side, three-particle
correlation has been proposed to look at the multi-particle
correlation in the emission pattern of particles. The di-hadron
correlation cannot distinguish different emission scenarios since
correlation only deals between emitted and the trigger particle.
However the three-particle correlation is capable of
distinguishing the different scenarios when simultaneous emission
of two particles are investigated with the trigger particle. If
the splitting structure of away-side is from large angle gluon
emission or deflection due to strong collective flow in an event,
the two associated particles will be clustered in a narrow cone on
a single-side of the away-jet direction. However, if the
production mechanism is Mach-cone shock wave or Cherenkov gluon
radiation, the partons in the shock-wave front or Cherenkov gluons
will be emitted conically around the away-side jet center in
single event. In this case, the two associated particles can  be
simultaneously on both sides of the $\Delta\phi = \phi_{assoc} -
\phi_{trig}$ distribution with respect to the opposite direction
of the trigger particle. Experimental studies of the
three-particle correlations have been reported by both the
STAR~\cite{fuqiang3p,Claude3p,star3p} and the
PHENIX~\cite{phenix3p} collaborators.

In our previous work, we reported observation of Mach-like
structure in di-hadron correlations from Au+Au collisions using a
multi-phase transport model (AMPT) where both partonic and
hadronic interactions are included~\cite{di-hadron}. Both parton
cascades and hadronic rescatterings can produce apparent di-hadron
correlations with Mach-like structures. But the hadronic
rescattering mechanism alone cannot reproduce the observed
experimental amplitude of Mach-like structure on the away-side,
which indicates that parton cascade processes are indispensable.
However, detailed dynamical mechanisms for the Mach-like structure
still await to be determined. In this Letter, we present a study
of three-particle correlation among one trigger particle and two
associated particles in Au + Au  collisions at $\sqrt{s_{NN}}$ =
200 GeV with the AMPT model. Three particle correlations in
regions of azimuthal angular directions of `cone', `deflected',
`center',`near' and `near-away', which will be defined later, will
be presented for Au + Au collisions from AMPT. With decreasing
number of participants, `center' correlations become more
dominant, and `cone' and `deflected' correlations seem to
disappear. Our results indicate that the three-particle
correlations in central collisions are mainly produced by partonic
Mach-like shock wave effect, while in peripheral collisions
deflected jet effect also contributes to the Mach-like structure.
Effects of hadronic rescatterings and parton cascades on
three-particle correlation are also investigated.

\begin{description}
   \item[II. Brief Description of the AMPT Model ]
\end{description}

AMPT model~\cite{AMPT}  is a hybrid model which consists of four main processes: the initial conditions,
partonic interactions, the conversion from partonic matter into hadronic matter and hadronic interactions. The
initial conditions, which include the spatial and momentum distributions of minijet partons and soft string
excitations, are obtained from the HIJING model~\cite{HIJING}. The excitation of strings will melt strings into
partons. Scatterings among partons are modelled by Zhang's parton cascade model  (ZPC) \cite{ZPC}, which at
present includes only two-body scatterings with cross section obtained from pQCD calculation with screening mass. In the
default version of AMPT model (we briefly call it as ``the default AMPT" model)~\cite{DAMPT}, partons are
recombined with their parent strings when they stop interactions, and the resulting strings are converted to
hadrons using the Lund string fragmentation model~\cite{Lund}. In  the string melting version of the AMPT model
(we briefly call it as ``the melting AMPT" model)\cite{SAMPT}, a quark coalescence model is used to
combine partons to form hadrons. Dynamics of the subsequent hadronic matter is then described by A Relativistic
Transport (ART) model \cite{ART}. Details of the AMPT model can be found in a recent review~\cite{AMPT}.
Previous studies \cite{AMPT,SAMPT,Jinhui} demonstrated that the partonic effect cannot be neglected and the melting
AMPT model is much more appropriate than the default AMPT model in describing nucleus-nucleus collisions at RHIC.
In the present work, the parton interaction cross section in the
AMPT model is assumed to  be 10 mb consistent with previous calculations~\cite{AMPT,Jinhui}.

\begin{description}
   \item[III. Analysis Method  ]
\end{description}

\begin{figure}
\includegraphics[scale=0.35]{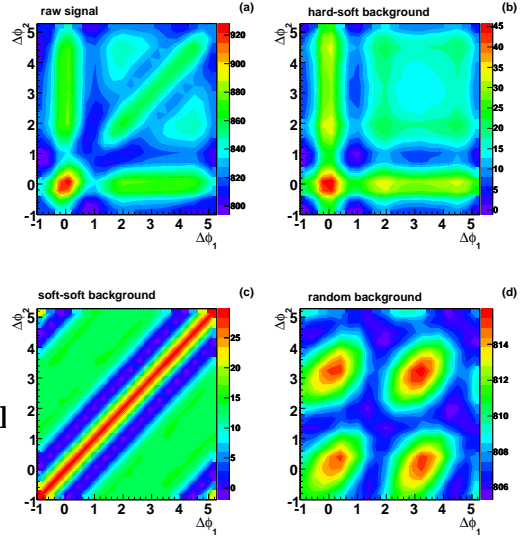}
\caption{\footnotesize Three-particle correlations in the top 10\%
 central Au+Au collisions at $\sqrt{s_{NN}}$ = 200 GeV from
the melting AMPT model with hadronic rescattering. (a): Raw
signal. (b): Hard-soft background. (c): Soft-soft background. (d):
Random background.
 }
\label{process}
\end{figure}

\begin{figure}
\includegraphics[scale=0.35]{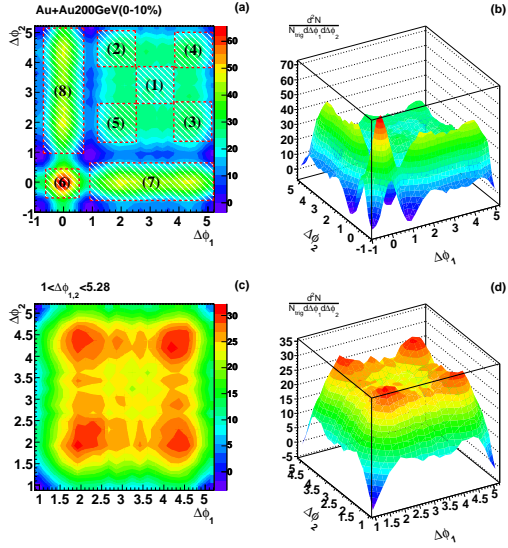}
\caption{\footnotesize  Background subtracted 3-particle
correlations in the top 10\% central Au+Au collisions at
$\sqrt{s_{NN}}$ = 200 GeV for the melting AMPT model with hadronic
rescattering. (a) and (b): Background subtracted 3-particle
correlations ($-1<\Delta\phi_{1,2}<5.28$). (c) and (d): background
subtracted 3-particle correlations ($1<\Delta\phi_{1,2}<5.28$)
from the selected regions (1,2,3,4,5). The azimuthal angular
regions are defined in panel (a) -- (1): `center' region; (2) and
(3): `cone' regions; (4) and (5): `deflected' regions; (6): `near'
region; (7) and (8): `near-away' regions.
 }
\label{signalAuAu}
\end{figure}

The mixing-event technique has been used in our three-particle
correlation analysis. The  $p_{T}$ window cuts for trigger and
associated particles were selected as  $2.5 < p_{T}^{trig} < 4$
GeV/$c$ and $1.0 < p_{T}^{assoc} < 2.5$ GeV/$c$, respectively.
Both trigger and associated particles were required to be within a
pseudo-rapidity window of $|\eta| < 1.0$, where $\eta$ is the
pseudo-rapidity of hadrons in the center-of-mass frame of Au+Au
collisions. In the same events, raw 3-particle correlation signals
in $\Delta\phi_{1} = \phi_{1} - \phi_{trig}$ versus
$\Delta\phi_{2} = \phi_{2} - \phi_{trig}$ were histogrammed.
Figure\ref{process}(a) shows the raw 3-particle correlation
distribution in the top 10\% central Au+Au collisions at
$\sqrt{s_{NN}}$ = 200 GeV in the melting AMPT model with hadronic
rescattering. Three classes of background contributions are
expected to contribute to the raw signal. The first one is the
hard-soft background which comes from a jet-induced
trigger-associated pair combined with a background associated
particle from bulk medium. We reproduced it by mixing a
trigger-associated pair with another associated particle from a
different event (Figure \ref{process}(b)). The second one is
soft-soft background which comes from an associated particle pair
combined with an uncorrelated trigger particle. We constructed
this background by mixing an associated particle pair from one
event with a trigger particle from a different event (Figure
\ref{process}(c)). The third one is a random combinatorial
background, which was produced by mixing a trigger particle and
two associated particles respectively from three different events
(Figure \ref{process}(d)). We required that the mixed events are
all from very close collision centralities which can be determined
by impact parameters in simulations. In order to subtract the
backgrounds from the raw signals, we set the signal at
$0.8<|\Delta\phi_{1,2}|<1.2$ to be zero. Figure~\ref{signalAuAu}
(a) and (b) give background subtracted 3-particle correlations in
the top 10\% central Au+Au collisions at $\sqrt{s_{NN}}$ = 200 GeV
in the melting AMPT model which includes hadronic rescattering. In
order to observe the 3-particle correlations among a trigger
particle and two away-side associated particles clearly, the
3-particle correlations in $1<\Delta\phi_{1,2}<5.28$ region are
shown with an expanded scale in Figure~\ref{signalAuAu} (c) and
(d).

\begin{description}
   \item[IV. Results and Discussions]
\end{description}

We divide the three-particle correlation distribution into several
regions based on the possible origin of the particle emission
pattern as shown in Figure 2a. The first one is `center' region
($|\Delta\phi_{1,2}-\pi|<0.5$) where the three-particle
correlation mainly comes from one trigger particle and two
associated particles in the center of away side. The `center'
correlations represent penetration ability of away-side jet. The
second one is `cone' region ($|\Delta\phi_{1}-(\pi\pm1)|<0.5$ and
$|\Delta\phi_{2}-(\pi\mp1)|<0.5$) where three-particle correlation
would form splitting peaks in di-hadron $\Delta\phi$ correlation
due to a conical emission pattern from away-side jet. It was
predicted that this conical emission may be produced by a
Mach-cone shock wave effect when a jet propagates faster than the
speed of sound in the medium creating shock wave front in the cone
region. The third one is `deflected' region
($|\Delta\phi_{1,2}-(\pi\pm1)|<0.5$) where associated particles
are emitted in the same side-ward region of the away-side jet in
one event. The `deflected' region three-particle correlations can
also yield splitting peaks on the away-side of two-particle
correlation distribution because though within one event the
away-side jet is deflected to one side only, but inclusively with
many events both sides of the jet direction can be populated. The
fourth region is the `near' area ($|\Delta\phi_{1,2}|<0.5$) where
three-particle correlation represents the correlation among
trigger particle and associated particles on near side of the
trigger direction. The fifth one is `near-away' correlation region
($1<\Delta\phi_{1,2}<5.28$ and $|\Delta\phi_{2,1}|<0.5$), which
reflects the correlation among trigger particle, one associated
particle on near side and another associated particle on away
side. The five regions have been marked with different numbers in
panel (a) of figure 2 for clarity. We will examine three-particle
correlations in the above five regions.

\begin{figure}
\includegraphics[scale=0.65]{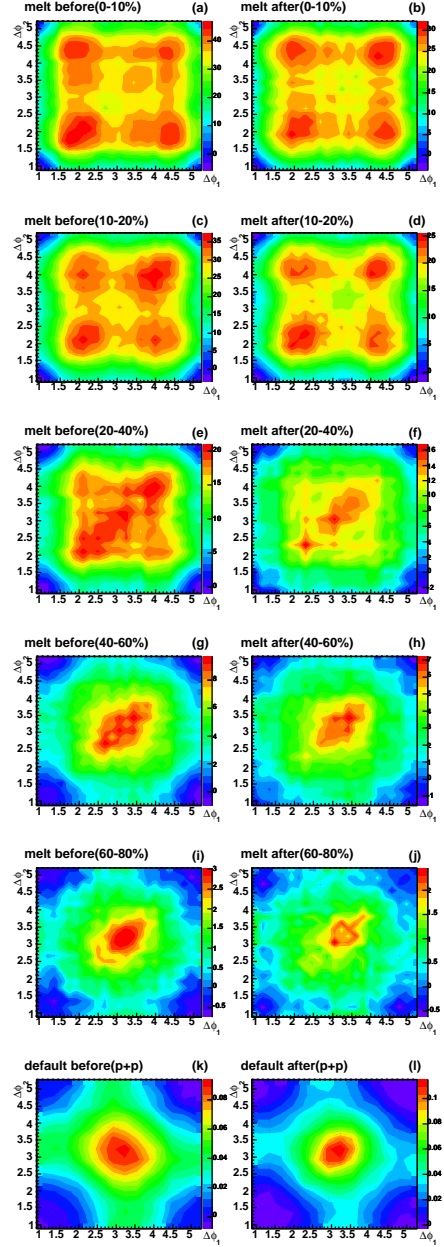}
\caption{\footnotesize  Background subtracted segmental 3-particle correlation areas ($1<\Delta\phi_{1,2}<5.28$)
in different centralities in Au+Au collisions at $\sqrt{s_{NN}}$ = 200 GeV in the melting AMPT model ((a)-(j)),
as well as p+p collisions at $\sqrt{s_{NN}}$ = 200 GeV in the default AMPT model ((k)-(l)). The left column from
(a) to (k) shows the results before hadronic rescattering (briefly named as ``melt before" or ``default before")
and the right column from (b) to (l) shows the results after hadronic rescattering (briefly named as ``melt
after" or ``default after"). (a) and (b): 0-10\%; (c) and (d): 10-20\%; (e) and (f): 20-40\%; (g) and (h):
40-60\%; (i) and (j): 60-80\%.  (k) and (l): p+p collisions.
 }
\label{areas}
\end{figure}

\begin{figure}
\includegraphics[scale=0.45]{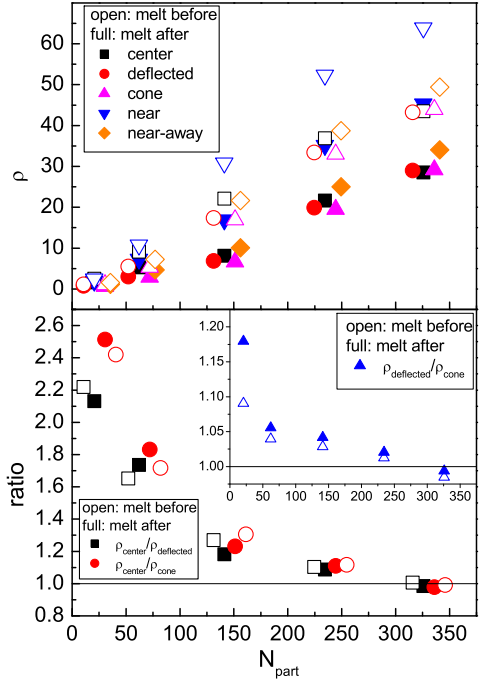}
\caption{\footnotesize The correlation density analysis for Au+Au
collisions at $\sqrt{s_{NN}}$ = 200 GeV in the melting AMPT model
before and after hadronic rescattering. Top panel: the average
three-particle correlation densities $\rho$ at different regions
as a function of $N_{part}$ ; Bottom panel: ratios of average
three-particle correlation density (`center'/`deflected' and
`center'/`cone') as a function of $N_{part}$;  The insert of the
bottom panel: ratio of average three-particle correlation density
(`deflected'/`cone') as a function of $N_{part}$. Note that some
points have been shifted slightly in $N_{part}$ axis for clarity.
} \label{ratio}
\end{figure}

Figure~\ref{areas} shows three-particle correlation distribution
in the ($1<\Delta\phi_{1,2}<5.28$) area from Au+Au collisions at
$\sqrt{s_{NN}}$ = 200 GeV with different centralities using the
melting AMPT model, and p+p collisions at $\sqrt{s_{NN}}$ = 200
GeV using the default AMPT model before and after hadronic
rescattering. Here we chose the default AMPT model to simulate p+p
collisions at $\sqrt{s_{NN}}$ = 200 GeV, since the string melting
mechanism has little effect on p+p collisions which has also been
demonstrated previously~\cite{AMPT}. Three-particle correlations
in all `center', `deflected' and `cone' regions can be observed in
central Au+Au collisions with the melting AMPT model regardless of
the inclusion of hadronic rescatterings. As the collisions become
more peripheral, the `deflected' and `cone' region correlations
gradually disappear until only the `center' correlations remain in
the most peripheral Au+Au collisions and p+p collisions.

In order to quantitatively express three-particle correlation strength in these different
regions, region-averaged three-particle correlation density $\rho$ is defined according to the following
equation:
\begin{equation}\label{density}
\rho=\frac{\int\int_{region}\frac{d^{2}N}
{N_{trig}d\Delta\phi_{1}d\Delta\phi_{2}}d\Delta\phi_{1}d\Delta\phi_{2}}{\int\int_{region}
d\Delta\phi_{1}\Delta\phi_{2}}.
\end{equation}
The top panel of Figure~\ref{ratio} shows three-particle
correlation densities $\rho$ in different regions as a function of
$N_{part}$ (number of participants) for Au+Au collisions at
$\sqrt{s_{NN}}$ = 200 GeV in the melting AMPT model before and
after hadronic rescattering.

Our results show that three-particle correlation densities decrease after hadronic rescattering process, which
indicates hadronic rescatterings could weaken three-particle correlation strength. However di-hadron
correlation is almost unchanged in this $p_{T}$ window selection in our previous work~\cite{di-hadron}. Such
a difference is indicative of enhanced sensitivity to hadronic rescatterings in the three-particle correlations in
comparison to the di-hadron correlations.

The bottom panel of Figure~\ref{ratio} shows the centrality
dependences of two ratios, namely the density ratios of
`center'/`deflected' and `center' /`cone', in Au+Au collisions at
$\sqrt{s_{NN}}$ = 200 GeV. Both ratios fall from above 2.0 in
peripheral collisions to near 1.0 in central collisions with the
increasing of $N_{part}$, which indicates that the strengths of
particle emission in the `cone' and in the `deflected' regions
increase dramatically in central collisions relative to the
particles in the `center' region. Since the `center' correlation
reflects the ability of $`punch-through'$ for the backward jet,
our results indicate that the backward jet can maintain the
original jet direction well in peripheral collisions while in
central collisions many particles are emitted in the `cone' and
the `deflected' directions, away from the original jet direction.

In the insert of Figure~\ref{ratio}, the ratio
of `deflected'/`cone' slightly decreases with $N_{part}$ and
approaches 1.0 in central collisions. The Mach-cone shock wave and the Chrenkov
gluon radiation scenarios predicted almost equal strength in the
three-particle correlations in the `deflected' ($\pi$$\pm$D,$\pi$$\pm$D) and  `cone'
($\pi$$\pm$D,$\pi$$\mp$D) regions, where D is the splitting parameter of away side
(i.e. half distance between two peaks on away side in di-hadron
$\Delta\phi$ correlation function). Our observed three-particle
correlations in the central Au+Au collisions from the AMPT model are consistent with
these model predictions. Such a consistency may be related to the hydrodynamic-like
behavior in the AMPT model due to strong parton-parton couplings and interactions.

More comments on the origin of the three-particle correlations in
the AMPT model are in order. The melting AMPT model was shown to
produce good descriptions of elliptic flow of identified hadrons
and even yielded the correct mass ordering of elliptic
flow~\cite{SAMPT,Jinhui}, which has been considered an important
feature of hydrodynamics models. Such an agreement can be
attributed to the large parton-parton interaction cross section in
the AMPT model, which leads to strong parton cascades that couples
partons together inducing the onset of hydrodynamical
behavior~\cite{Zhang99}. However, in another hydrodynamic
model~\cite{Chaudhuri} the signal of Mach-cone shock waves can
hardly be observed in the di-hadron correlations. It appears that
the large strength of parton cascades and coupling of partons as
described in the AMPT model bring about the conic emission pattern
on the away-side  prominently. The linearized hydrodynamical
approximation may not be adequate for the strong jet-medium
interaction region where the medium also experiences rapid
variation of energy density and without sufficient
thermalization~\cite{Casalderrey}. On the other hand, the observed
three-particle correlations may partly stem from deflected jets
(represented by $\frac{\rho_{deflected}}{\rho_{cone}} - 1$ )  in
peripheral collisions where `center' correlation  becomes
dominated. In the AMPT model there is no inclusion of large angle
gluon bremsstrahlung mechanism~\cite{Vitev} which may also play a
role in real collisions. In addition, we note that the backward
jet may also be distributed over a wide rapidity
range~\cite{Vitev,RenkRapidityMach} beyond our narrow $\eta$
window cut. In our model, we used LO pQCD cross sections from
HIJING model for the minijet production, which has successfully
described the suppression of back-to-back
jets~\cite{XinNianPLBdAu}. Our selection of the $\eta$ window cut
was to match the detector acceptance of the RHIC experiments.

\begin{figure}
\includegraphics[scale=0.35]{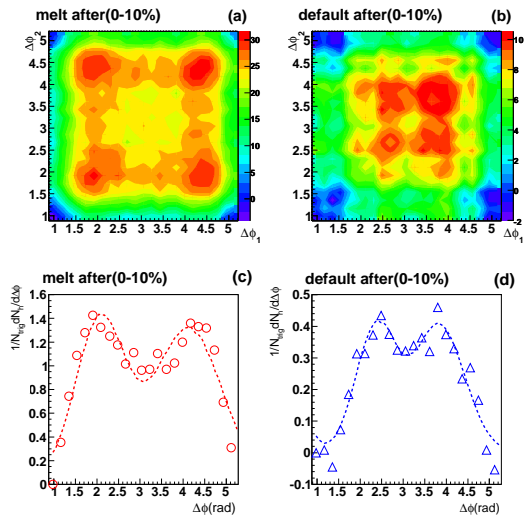}
\caption{\footnotesize  Background subtracted three-particle
correlations in selected ($1<\Delta\phi_{1,2}<5.28$) regions ((a)
and (b)) and away-side di-hadron correlations ((c) and (d)) in the
top 10\% Au+Au collisions at $\sqrt{s_{NN}}$ = 200 GeV in the
melting AMPT model (left column) and the default AMPT model (right
column) after hadronic rescattering.
 }
\label{2areas}
\end{figure}

In addition, we studied the effect of parton cascades on
three-particle and di-hadron correlation by comparing the results
of the default AMPT model and the melting AMPT model.
Figure~\ref{2areas}(a) and (b) give three-particle correlations in
selected ($1<\Delta\phi_{1,2}<5.28$) regions for the melting AMPT
model and the default AMPT model. Note both cases are the results
after hadronic rescattering in the top 10\% Au+Au collisions at
$\sqrt{s_{NN}}$ = 200 GeV. Though with large statistical errors,
the default AMPT model seems to produce a three-particle
correlation, but the three-particle correlation area is
considerably less than that from the melting AMPT model. It is
consistent with the results of di-hadron correlation in our
previous work (see Figure~\ref{2areas}(c) and (d)) that concluded
that hadronic rescattering alone cannot reproduce a splitting
parameter of Mach-like structure on away side large enough to
match the experimental measurements~\cite{di-hadron}.

\begin{figure}
\includegraphics[scale=0.37]{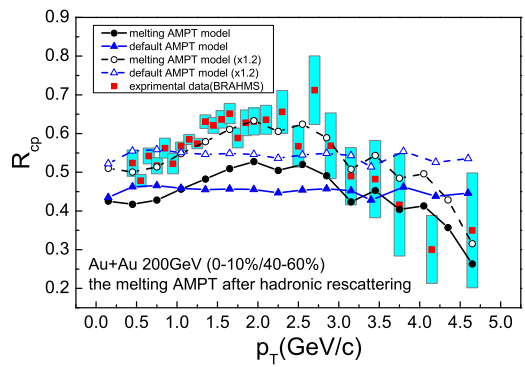}
\caption{\footnotesize  The $p_{T}$ dependences of nuclear modification factor $R_{cp}$ of charge hadrons for
0-10\%/40-60\% in Au+Au collisions at $\sqrt{s_{NN}}$ = 200 GeV in the melting and default AMPT model with
hadronic rescattering. The experimental data come from Ref.~\cite{rcp}.
 }
\label{rcp}
\end{figure}

The nuclear modification factor, $R_{cp}$, is also considered a useful probe of the
energy loss of high $p_{T}$ partons in the dense medium created in nucleus-nucleus collisions.
Figure~\ref{rcp} shows the transverse momentum dependences of nuclear
modification factor $R_{cp}$ of charge hadrons in the melting and default AMPT model with hadronic rescattering.
The $R_{cp}$ is difined by following formula:
\\
$R_{cp}=\frac{N_{bin}|_P}{N_{bin}|_C} \times
\frac{\frac{d^{2}N}{p_{T}dp_{T}d\eta}|_{C}}{\frac{d^{2}N}{p_{T}dp_{T}d\eta}|_{P}}$,
\\
where the \textbf{C}entral and the \textbf{P}eripheral collision centralities are 0-10\% and 40-60\%, and the
respective number of binary collisions $N_{bin}$= 939.4 (0-10\%), 93.7 (40-60\%).
The $R_{cp}$ in the melting AMPT model is of similar shape of experimental data, which can match
experimental data well if scaled by a factor 1.2. However the $R_{cp}$ from the default AMPT model seems to be
independent of $p_{T}$ and inconsistent with experimental data.
The partonic interactions in the melting AMPT model appear
essential to describe the shape of nuclear modification factor as a function of $p_T$ in Au+Au
collisions. Furthermore, $R_{cp}$ is suppressed more heavily in higher $p_{T}$ range ($p_{T}$ $>$ 3.5 GeV/c) in
the melting AMPT model than in the default AMPT model, which may indicate that more energies are lost into the
medium by parton cascade mechanism especially for high $p_{T}$ particles, which is expected to be in favor of
the formation of partonic Mach-like shock waves.

\begin{description}
   \item[V. Conclusions ]
\end{description}

Three-particle correlations have been extracted by using
event-mixing technique in a multi-phase transport model with both
partonic and hadronic interactions. Correlations in different
azimuthal angular regions with respect to the trigger jet
direction, so called `center' ,`deflected', `cone', `near' and
`near-away', have been discussed for Au+Au collisions  at
$\sqrt{s_{NN}}$ = 200 GeV. The AMPT results with and without
hadronic rescattering are also compared. The `center'
three-particle correlation becomes more and more dominant with the
decreasing of number of participants, which may reflect the
centrality dependence of partonic density and the strength of
partonic interactions. The density ratio of `deflected'/`cone'
approaching 1.0 in central collisions indicates that the
three-particle correlation in central collisions is mainly
produced by a partonic Mach-like shock wave mechanism, and in
peripheral collisions deflected jet mechanism also contributes.
The partonic Mach-like shock wave mainly originates from strong
partonic interactions in dense partonic matter. The three-particle
correlations are also sensitive to hadronic rescatterings,
therefore the effect of hadronic rescattering may need to be
considered in quantitative studies. The default AMPT model, where
only the hadronic rescattering mechanism plays a dominant role,
produces a three-particle correlation area much smaller than the
melting AMPT model which includes both parton cascade and hadron
rescattering mechanisms. Our AMPT calculation of three-particle
correlations re-affirms our previous conclusion from di-hadron
correlation studies that hadronic rescattering alone cannot
produce an amplitude of Mach-like cone on away side large enough
to match the experimental data. Parton cascade mechanism is
essential and important in order to describe the amplitude of
observed experimental Mach-like structure.

\begin{description}
   \item[Acknowledgements ]
\end{description}

This work was supported in part by the National Natural Science
Foundation of China  under Grant No. 10610285 and 10328259, the
Knowledge Innovation Project of Chinese Academy of Sciences under
Grant No. KJCX2-YW-A14 and KJXC3-SYW-N2, and  the Shanghai
Development Foundation for Science and Technology under Grant
Numbers 05XD14021 and 06JC14082.


\end{document}